\documentclass[11pt,letterpaper]{article}
\usepackage[T1]{fontenc}
\usepackage[utf8]{inputenc}
\usepackage{lmodern}
\usepackage[margin=1in]{geometry}
\usepackage{amsmath,amssymb,booktabs,array,microtype}
\usepackage{xcolor}
\usepackage{pgfplots}
\pgfplotsset{compat=1.18}
\usepackage[hidelinks]{hyperref}
\usepackage[round,authoryear]{natbib}
\newcommand{\R}{\mathbb{R}}
\newcommand{\E}{\mathbb{E}}
\title{Tail Instability and Scale Control in Conditional Normalizing Flows for Treasury Yield-Curve Scenarios}
\author{Arin Mohanty}
\date{October 9, 2026}
\begin{document}
\maketitle
\begin{abstract}
A conditional normalizing flow can produce extreme scenarios even when its point forecasts appear reasonable. We compare two fitted models for daily Treasury yield changes at eleven maturities reconstructed from Federal Reserve curve parameters. Both use an LSTM, an auxiliary location head, and an affine residual flow. On a previously unused 60-date window, the frozen Student-$t$ model generated an absolute standardized coordinate of $3.43\times10^7$. A Gaussian-base model with tighter coupling scale control reached 14.72. Each generated 600,000 vectors. The original model's scenario-mean MSE was 33,908.95, while its location-head MSE was 0.3376. The stabilized model's two estimates remained close. Across five development windows, the stabilized maximum was 17.33 among three million vectors. Stability did not imply better prediction: historical simulation had better distributional scores and coverage closer to nominal on the new window. The original flow also had a better log score despite its extreme samples. Frozen-model evaluation is reproducible, but the comparison does not isolate architectural effects or establish a general forecasting advantage.
\end{abstract}

\section{Introduction}
Joint yield-curve scenarios describe how rates at different maturities may move together. Their use in risk measurement requires more than accurate point forecasts. A model may fit typical changes yet generate rare values large enough to distort Monte Carlo means, tail averages, and valuations.

Normalizing flows use invertible transformations to provide tractable densities and sampling \citep{dinh,papamakarios}. Conditioning on history lets the distribution change over time \citep{winkler}. Invertibility alone does not ensure numerical stability. \citet{behrmann} study exploding inverses and ill-conditioning, while \citet{andrade} proposes scale controls for excessive sample magnitudes in variational inference.

We study a conditional residual flow for Treasury yield changes. The original Student-$t$ model generated extreme values despite coupling log-scales bounded at $\pm3$. A modified model combines a Gaussian base with a tighter smooth bound of $\pm0.5$. Its sampled extremes were much smaller, but predictive performance remained mixed against simple alternatives.

Flow-tail theory relates attainable tail behavior to the base distribution and transport regularity \citep{jaini,laszkiewicz}. Preventing numerical extremes and fitting rare market changes are separate tasks.

The five development windows informed the stabilization choices, with one saved fitted model per specification and fold. We also evaluate a subsequent window using frozen checkpoints and a local plan described in Section~\ref{sec:confirmation}. The checkpoints support evaluation replay. An initialization-order issue limits reproduction of the original training runs.

\section{Data and evaluation design}
\subsection{Reconstructed yield panel}
The main experiments use the Federal Reserve Board's G\"urkaynak--Sack--Wright fitted-curve parameters \citep{gsw}. We evaluate a saved snapshot at 1, 3, and 6 months and 1, 2, 3, 5, 7, 10, 20, and 30 years. The panel contains 16,170 dates from June 14, 1961 through April 10, 2026. These values are fitted zero-coupon yields. The code uses DGS-style column names, but the values are not FRED constant-maturity observations.

The reconstruction uses the Svensson form, reducing to its Nelson--Siegel component where the second curvature term is unavailable. This follows the Federal Reserve's distinction between pre-1980 Nelson--Siegel estimates and later Svensson estimates. The source data can be revised \citep{feddata}; we use one saved vintage rather than historical real-time vintages.

Sub-year values extrapolate below the fitted curve's supported maturity range. Early long-maturity values can also extend beyond the maturities supporting the fit. The eleven reconstructed coordinates therefore differ in reliability. This matters because many of the original model's largest samples occurred at the 20-year coordinate.

\subsection{Forecast targets and temporal splits}
Let $y_t\in\R^{11}$ denote the fitted yields, measured in percentage points, and let $\Delta y_t=y_t-y_{t-1}$. For each fold, the training mean $\mu_j$ and standard deviation $\sigma_j$ are computed from changes strictly before the first evaluation date. The standardized target is
\begin{equation}
 x_{t,j}=\frac{\Delta y_{t,j}-\mu_j}{\sigma_j},\qquad j=1,\ldots,11.
\end{equation}
Conditional models use the preceding 60 observed change vectors. Model fitting and scaler estimation use only pre-window data. During evaluation, histories update with observations available at each forecast origin. The predictions are one step ahead, not a single 60-step path.

\begin{table}[htbp]
\centering\small
\caption{Evaluation windows and available training sequences. Each fold contains 60 forecast dates. Training sequences contain 60 lagged observations.}
\begin{tabular}{cllr}
\toprule Fold & First target date & Last target date & Training sequences\\\midrule
1 & 2002-04-03 & 2002-06-26 & 10,098\\
2 & 2008-03-12 & 2008-06-04 & 11,585\\
3 & 2014-02-21 & 2014-05-16 & 13,074\\
4 & 2020-02-03 & 2020-04-28 & 14,561\\
5 & 2026-01-15 & 2026-04-10 & 16,049\\\bottomrule
\end{tabular}
\label{tab:folds}
\end{table}

The windows include the early financial crisis and the onset of COVID-19. They cover 300 selected dates, not the full period from 2002--2026. Stabilization choices used these regimes, including variant screens on folds 1 and 2. Pre-window fitting prevents direct temporal leakage but does not remove this selection dependence.

\section{Conditional residual flow and stabilization}
\subsection{Architecture and training}
A two-layer LSTM with hidden dimension 64 encodes the input history. Its final hidden state feeds a location head $m_\theta$ and a 24-dimensional context projection $c_\theta$:
\begin{equation}
 h_t=\operatorname{LSTM}_\theta(x_{t-60:t-1}),\qquad
 m_t=m_\theta(h_t),\qquad c_t=c_\theta(h_t).
\end{equation}
The location head has a 64-unit hidden layer with SiLU activation. The context projection uses Tanh and dropout. Eight affine coupling layers transform the residual $r_t=x_t-m_t$ into $z_t=f_\phi(r_t;c_t)$. Masks alternate, and ActNorm precedes each layer. Every other coupling layer receives context through a sigmoid gate; the others are unconditional.

The log density and sampling equations are
\begin{align}
 \log p(x_t\mid h_t)&=\log p_Z\!\left(f_\phi(x_t-m_t;c_t)\right)
   +\log\left|\det J_{f_\phi}(x_t-m_t;c_t)\right|,\\
 x_t^{(s)}&=m_t+f_\phi^{-1}(z_t^{(s)};c_t),\qquad z_t^{(s)}\sim p_Z.
\end{align}
Training optimizes the location head and encoder with MSE for 50 epochs. Another 500 epochs minimize flow negative log likelihood plus $0.1$ times the location-head MSE. Appendix~\ref{app:config} gives the configuration.

The residuals are not constrained to have zero mean, so
\begin{equation}
 \E[x_t\mid h_t]=m_t+\E[f_\phi^{-1}(Z;c_t)\mid h_t]
\end{equation}
need not equal $m_t$. We therefore report the location-head MSE separately from the Monte Carlo scenario-mean MSE.

\subsection{Failure mechanism and modified specification}
For a coupling layer with frozen coordinates $a$ and transformed coordinates $b$, the inverse operation has the form
\begin{equation}
 r_b=\{u_b-t(u_a,c)\}\odot\exp\{-s(u_a,c)\}.
\end{equation}
The original flow used a learnable Student-$t$ base and log-scales clamped to $[-3,3]$. One inverse coupling can still multiply a coordinate by up to $e^3\simeq20.09$. A layer-level audit reproduced the extremes and traced contributions from large latent draws, inverse scale factors, and translations.

The stabilized specification uses a standard Gaussian base and replaces the coupling scale rule with
\begin{equation}
 s=0.5\tanh(s_{\rm raw}/0.5).
 \label{eq:bound}
\end{equation}
The direct coupling multiplier then lies between $e^{-0.5}$ and $e^{0.5}$. Translations and ActNorm parameters remain unrestricted, and the Gaussian base has unbounded support. Equation~\eqref{eq:bound} limits one source of amplification. It does not bound the full output or guarantee a globally Lipschitz network. The nonlinear transformation also means that a Gaussian base need not produce Gaussian yields.

The modification applies established stabilization ideas to this model and dataset \citep{behrmann,andrade}. Because the base and scale bound change together, the comparison cannot separate their effects.

\section{Benchmarks and metrics}
\subsection{Implemented comparators}
The point baseline predicts zero standardized change, which corresponds to the \emph{training mean raw change}. The saved code calls it ``Zero\_Change.'' We also reconstruct a literal zero raw-change baseline using the same panel and pre-fold scalers. Its standardized prediction is $-\mu/\sigma$.

The probabilistic comparators are an expanding-history multivariate Gaussian, historical simulation, Gaussian VAR, and PCA-VAR. Historical simulation uses a fixed pool of 252 pre-fold changes throughout each window. Training-only BIC selects VAR lags from 1--5, and PCA retains at least 95\% of training variance. The Gaussian uses the full eligible pre-fold history.

The neural comparators are an unconditional RealNVP and an LSTM-conditional RealNVP. They differ in layer counts, scale controls, normalization, training, and use of a location head. A dynamic Nelson--Siegel baseline was not completed under this protocol and is omitted.

\subsection{Point, distribution, and tail diagnostics}
For $T$ forecast dates and $d=11$ maturities, standardized MSE is
\begin{equation}
 \operatorname{MSE}=\frac{1}{Td}\sum_{t,j}(\hat x_{t,j}-x_{t,j})^2.
\end{equation}
Raw RMSE reverses standardization, converts errors to basis points, and pools squared errors before taking the square root. Standardized MSE weights errors by each fold's training volatility, so its ranking can differ from raw RMSE.

The central 90\% marginal interval spans the 5th to 95th sample percentiles at each date and maturity. Coverage averages the resulting indicators; it is not simultaneous full-curve coverage. Widths use standardized units.

For predictive distribution $F$ and realized standardized vector $x$, the Energy Score is
\begin{equation}
 \operatorname{ES}(F,x)=\E\|X-x\|_2-\frac12\E\|X-X'\|_2,
 \qquad X,X'\stackrel{\text{iid}}{\sim}F.
\end{equation}
Lower values are better \citep{gneiting}. The implementation estimates Energy Score from 1,000 scenarios and their pairwise distances. Table~\ref{tab:distribution} and the paired comparisons use the mean of saved \emph{per-date} scores. Aggregate files use different random subsamples and give slightly different estimates (Appendix~\ref{app:provenance}).

Tail diagnostics report the largest absolute standardized coordinate, non-finite outputs, and the share exceeding diagnostic thresholds. These sample checks do not estimate a specified market stress probability. A saved correlation-Frobenius metric compares within-day predictive correlations with correlations across realized dates. We exclude it from rankings because those quantities differ.

\section{Results}
\subsection{Observed tail stability}
The original model's stress audit reached an absolute standardized coordinate of about $7.07\times10^{10}$, using up to 50,000 samples per forecast origin. Of twenty top extremes recorded in walk-forward runs, eighteen occurred at the reconstructed 20-year coordinate. The inverse-layer audit reproduced them, ruling out de-standardization as their sole source.

The stabilized model generated 10,000 vectors on each of 300 dates, for three million vectors. The largest absolute coordinate was 17.33. None exceeded 20, and none were NaN or infinite. Realized targets under the same scalers had fold maxima of 2.54, 4.03, 2.78, 5.25, and 2.52 (Appendix~\ref{app:config}). The threshold of 20 was a development diagnostic, not a market quantile. The two models' development maxima also use different protocols; Section~\ref{sec:confirmation} compares them under equal counts.

\begin{figure}[htbp]
\centering
\begin{tikzpicture}
\begin{axis}[width=.92\linewidth,height=5.3cm,xmin=.5,xmax=5.5,ymin=0,ymax=22,
xtick={1,2,3,4,5},xticklabels={2002,2008,2014,2020,2026},
ylabel={Maximum absolute standardized move},xlabel={Evaluation window},
grid=major,grid style={gray!15},ybar,bar width=22pt,enlarge x limits=false]
\addplot[fill=blue!45,draw=blue!70!black] coordinates {(1,15.72258) (2,16.12618) (3,13.23536) (4,17.32796) (5,16.97145)};
\addplot[red!70!black,dashed,sharp plot,update limits=false] coordinates {(.5,20) (5.5,20)};
\end{axis}
\end{tikzpicture}
\caption{Finite-sample tail diagnostic for the stabilized flow. The dashed line is the diagnostic threshold of 20 standardized units. Each bar summarizes 600,000 scenario vectors.}
\label{fig:tails}
\end{figure}
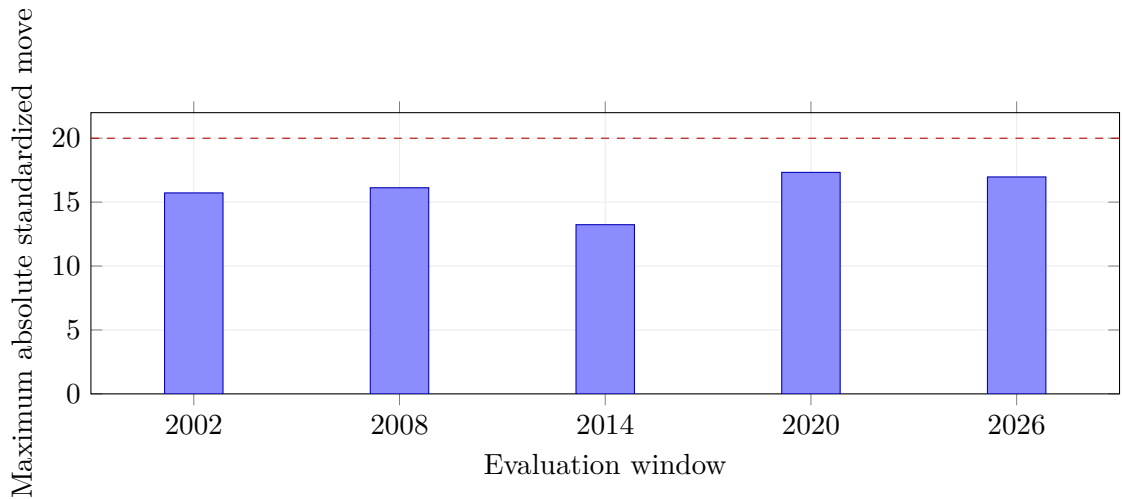

Variant screens on folds 1 and 2 provide context. Removing the scale bound produced a non-finite training loss. A Student-$t$ variant with the tighter bound and a Gaussian variant with a looser bound both produced larger extremes. Training lengths, initialization, and sample counts were not consistently matched, so these screens do not isolate either component's effect.

\subsection{Point forecasting}
Table~\ref{tab:points} separates the location head from scenario means. The stabilized head reduces standardized MSE against the training-mean baseline by about 0.0061, or 0.97\%. Its scenario mean has higher MSE than either simple baseline. Raw RMSE favors the simple baselines over the stabilized head.

\begin{table}[htbp]
\centering\small
\caption{Point forecasting across the five equal-sized windows. Raw RMSE is marked n/a where comparable raw-error series are unavailable. The literal zero-change row was recomputed from the saved panel.}
\begin{tabular}{lrr}
\toprule Estimator & Standardized MSE & Raw RMSE (bp)\\\midrule
Stabilized flow: location head & 0.6223 & 6.320\\
Original residual flow: location head & 0.6265 & n/a\\
Training-mean change / expanding Gaussian mean & 0.6284 & 6.279\\
Literal zero raw change & 0.6285 & 6.279\\
Stabilized flow: scenario mean & 0.6312 & n/a\\
Fixed-pool historical simulation: scenario mean & 0.6321 & 6.296\\
Unconditional RealNVP: scenario mean & 0.6441 & n/a\\
PCA-VAR: conditional mean & 0.6520 & 6.410\\
Gaussian VAR: conditional mean & 0.6607 & 6.460\\
LSTM-conditional RealNVP: scenario mean & 0.6738 & 6.487\\\bottomrule
\end{tabular}
\label{tab:points}
\end{table}

We compute paired circular block-bootstrap intervals for daily loss differences. Each replicate resamples blocks within each window and weights all five windows equally. With 20,000 replicates and block length 10, the head-minus-training-mean difference is $-0.0061$, with a 95\% percentile interval of $[-0.0176,0.0035]$. Block lengths 5 and 20 also give intervals containing zero. Against literal zero change, the difference is $-0.0062$, with interval $[-0.0179,0.0035]$.

These descriptive intervals condition on the fitted models and selected regimes. They exclude model selection, refitting, training variability, and dependence between regimes.

\subsection{Distributional scores and coverage}
Historical simulation has the lowest Energy Score in Table~\ref{tab:distribution}. LSTM-conditional RealNVP has the narrowest intervals (1.898 versus 1.952 for historical simulation). Both have coverage below nominal. The stabilized flow has wider intervals and higher coverage; the Gaussian and VAR models are wider still.

\begin{table}[htbp]
\centering\small
\caption{Distributional evaluation from saved per-date metrics. Coverage is marginal; widths and Energy Scores use standardized changes. The final column is the maximum saved absolute standardized coordinate.}
\begin{tabular}{lrrrr}
\toprule Model & Coverage (\%) & Width & Energy & Maximum\\\midrule
Stabilized residual flow & 92.85 & 2.456 & 1.5112 & 17.33\\
Expanding Gaussian & 94.24 & 3.290 & 1.6268 & 5.55\\
Fixed-pool historical simulation & 85.64 & 1.952 & 1.5040 & 7.50\\
Gaussian VAR & 93.82 & 3.208 & 1.6503 & 5.736\\
PCA-VAR & 93.36 & 3.154 & 1.6347 & 5.737\\
LSTM-conditional RealNVP & 83.97 & 1.898 & 1.5134 & 1502.52\\\bottomrule
\end{tabular}
\label{tab:distribution}
\end{table}

The stabilized model's coverage varies across folds: 95.45\%, 87.12\%, 99.39\%, 85.61\%, and 96.67\%. Coverage falls below 90\% in the stress windows despite exceeding it on average.

The paired block bootstrap gives a stabilized-minus-historical Energy Score difference of 0.0072, with a block-10 interval of $[-0.0305,0.0406]$. The point estimate favors historical simulation, but the interval includes zero.

\subsection{Subsequent window with frozen checkpoints}\label{sec:confirmation}
Before downloading the new snapshot or running this evaluation, we recorded a local plan. It selected the first 60 valid target dates after April 10, 2026 and retained the fold-5 checkpoints and scalers. It specified 10,000 vectors per model per date with no further tuning. The October 9, 2026 Federal Reserve snapshot yields a window from April 13 to July 8, 2026. Both neural models were trained on data before January 15, 2026. We reconstruct the targets and their preceding 60-date contexts from the new snapshot. Levels on the preceding 61 dates differ from the old snapshot by at most $5.15\times10^{-11}$ basis points.

Historical simulation uses the 252 changes immediately preceding April 13, standardized with the frozen fold-5 scaler. Its pool is more recent than the neural training data, although neural contexts also update with new observations. This comparison therefore does not match refitting frequency. The plan is saved in \texttt{confirmation\_protocol.json}, with a SHA-256 hash in the run manifest. It has no external registration or public timestamped commit. Local timestamps agree with the reported sequence but cannot independently verify it; this is not a publicly preregistered experiment.

\begin{table}[htbp]
\centering\small
\caption{Previously unused 60-date window, April 13--July 8, 2026. Every probabilistic model generates 600,000 vectors. All generated coordinates were finite. NLL is average joint negative log likelihood in standardized coordinates; n/a denotes an inapplicable metric, including a continuous density for historical simulation.}
\begin{tabular}{lrrr}
\toprule Metric & Original flow & Stabilized flow & HistSim-252\\\midrule
Maximum absolute coordinate & $3.43\times10^7$ & 14.72 & 3.71\\
Coordinate share above 20 & $9.79\times10^{-5}$ & 0 & 0\\
Location-head MSE & 0.3376 & 0.3355 & n/a\\
Scenario-mean MSE & 33,908.95 & 0.3351 & 0.3376\\
90\% marginal coverage (\%) & 95.91 & 96.82 & 91.52\\
Average standardized width & 2.101 & 2.326 & 1.712\\
Energy Score & 1.1613 & 1.1645 & 1.1348\\
Mean marginal CRPS & 0.3022 & 0.2981 & 0.2927\\
Joint NLL & $-38.77$ & $-37.42$ & n/a\\\bottomrule
\end{tabular}
\label{tab:confirmation}
\end{table}

The original flow's largest coordinate occurs at the extrapolated three-month maturity. Its three-year coordinate also reaches $5.97\times10^6$, so the failure extends to the supported maturity range. Extreme samples distort its scenario mean while leaving its location-head MSE near 0.3376. The stabilized flow stays below the diagnostic threshold, and its scenario-mean and head MSEs remain close. The training-mean and literal zero-change baselines achieve MSEs of 0.33730 and 0.33735, respectively.

Marginal CRPS uses all 10,000 scenarios and the empirical absolute-distance formula, averaged across maturities and dates. Energy Score uses the planned 1,000-scenario subsample. Historical simulation has the lowest values of both scores and coverage closest to 90\%. The original flow has lower NLL than the stabilized flow despite its extreme samples. Average log score did not flag the tail failure that distorted its Monte Carlo mean. Negative NLLs are possible because densities can exceed one.

No models were retrained or tuned, and forecast outcomes did not determine the window.

\subsection{Exploratory risk diagnostic}
A separate constant-maturity exercise shows why breach counts alone can be misleading. The frozen model's 95\% VaR breach count is close to nominal, but the breaches cluster. Appendix~\ref{app:risk} gives the checkpoints, valuation proxy, breach dates, and tests.

\section{Limitations}

The reconstructed panel limits financial interpretation. Some maturities are extrapolated, and a single data vintage cannot recreate the information available to historical forecasters. Eleven coordinates also come from a low-dimensional fitted curve, which complicates the interpretation of joint density scores. Broader market claims require evaluation on supported maturities or observed constant-maturity data.

The five historical windows informed model development. The subsequent evaluation adds unseen targets, but only one window, and its numerical threshold was chosen during development. One fitted model per fold and the lack of matched ablations prevent separating architectural effects or measuring training variability.

The original training runner constructs the model before setting its recorded seed. Seed labels therefore do not fully specify initialization. Checkpoint evaluation is reproducible, but deterministic training reproduction has not been established.

Predictive evaluation is incomplete. Held-out log scores and marginal CRPS cover only the subsequent window. Variogram scores and formal conditional calibration tests are absent. Limiting numerical extremes could reduce the fit to rare changes; flow-tail theory highlights the roles of base distributions and transport regularity \citep{jaini,laszkiewicz}. The observed 90\% undercoverage does not identify its cause. Sample maxima and the threshold of 20 also do not assess 99\% or 99.9\% quantile accuracy. Stronger tail claims require longer evaluations and matched comparisons across training seeds.

The risk exercise is short, overlaps development data, and uses a constant-maturity price proxy. Its coverage and clustering results do not validate a traded-instrument model or an operational risk system.

\section{Conclusion}
The original Student-$t$ flow generated extreme Treasury yield scenarios despite bounded coupling scales. A Gaussian base with tighter scale control avoided these failures in the completed evaluations. On a new 60-date window, frozen checkpoints reproduced the contrast under equal sample counts. The original model's Monte Carlo mean was distorted even though its location head and log score appeared reasonable. Historical simulation still achieved the best distributional scores on that window. This case study shows why flow evaluation should distinguish point forecasts, scenario means, sampled tails, and calibration. The observed stabilization applies to these fitted models; its causes and broader tail accuracy remain unresolved.

\section*{Acknowledgments}
I thank Aaron Y. S. Kim for research supervision. The interpretations and remaining errors are my responsibility.

\section*{Data and computational provenance}
The analysis uses public Federal Reserve fitted-curve data and FRED Treasury yields. The arXiv source package includes reproducibility files under \texttt{anc/reproduction/}, accessible through its ancillary files or source download. The bundle contains source-file hashes, audit scripts, and installation instructions. It also includes the protocol, code, frozen checkpoints, scalers, and input snapshot for replaying the subsequent-window evaluation. The audits check the 300 development dates, baseline scalers, pooled errors, bootstrap intervals, and risk-test arithmetic. Checkpoints and scalers were unchanged during evaluation.

\appendix
\section{Model configuration and fold results}\label{app:config}
The stabilized model uses a 60-date history and a two-layer, 64-unit LSTM with dropout 0.15. Its location head is a $64\to64\to11$ SiLU network. The $64\to24$ Tanh context projection has dropout 0.20. Eight coupling layers alternate masks; every other layer is conditional. Scale and translation networks each have two 128-unit SiLU hidden layers. ActNorm precedes every coupling. The base is standard Gaussian, coupling log-scales follow Equation~\eqref{eq:bound}, and translations are not clipped.

The training code uses batches of 64 and Adam with learning rate $10^{-3}$ during warm-up. Joint training uses AdamW with encoder learning rate $3\times10^{-4}$, flow learning rate $10^{-3}$, and weight decay $10^{-4}$. A cosine schedule spans the 500 joint epochs, and gradients are clipped at norm 1. Fold seed labels are 43--47, subject to the initialization limitation in Section 6.

\begin{table}[htbp]
\centering\small
\caption{Stabilized-model fold results and realized targets. ``Head'' and ``Mean'' are distinct point estimators. Both maxima are absolute standardized coordinates; generated maxima summarize 600,000 vectors per fold, whereas realized maxima summarize 60 vectors.}
\begin{tabular}{crrrrr}
\toprule Fold & Head MSE & Mean MSE & Coverage (\%) & Generated max & Realized max\\\midrule
1 & 0.4192 & 0.4584 & 95.45 & 15.72 & 2.54\\
2 & 1.1179 & 1.1009 & 87.12 & 16.13 & 4.03\\
3 & 0.1717 & 0.1722 & 99.39 & 13.24 & 2.78\\
4 & 1.0644 & 1.0861 & 85.61 & 17.33 & 5.25\\
5 & 0.3382 & 0.3385 & 96.67 & 16.97 & 2.52\\\bottomrule
\end{tabular}
\end{table}

\section{Saved-output reconciliation}\label{app:provenance}
The audit script reads the original CSVs without changing them and records SHA-256 hashes. The evidence directory contains aggregate and daily metrics, the reconstructed panel, tail summaries, extreme-sample diagnostics, and risk tests. Stabilized checkpoints for folds 1--2 come from the strict-screen directory; folds 3--5 come from the completed five-fold directory.

Point MSE and raw RMSE agree between the daily files and aggregates. Energy Scores differ because the scripts use separate random subsamples for fold-level and daily estimates. Table~\ref{tab:distribution} uses daily-score means. For the stabilized flow and historical simulation, aggregate files report 1.5083 and 1.4976, while daily-score means are 1.5112 and 1.5040.

The paired bootstrap uses seed 20261009 and 20,000 replicates. Within each fold, it draws block starts uniformly from the 60 dates, joins circular blocks, and truncates to 60 observations. The same indices resample each pair's loss difference. The companion CSV retains 2.5th and 97.5th percentiles for block lengths 5, 10, and 20.

\section{Exploratory constant-maturity risk proxy}\label{app:risk}
The stabilized architecture is trained on eleven FRED constant-maturity series in \nolinkurl{treasury_yields_daily.csv}. Scalers use pre-window data. The checkpoint \texttt{stable\_clip0p5\_frozen.pt} in the risk-backtest directory was trained through April 28, 2025, before the first forecast origin on April 29. It is separate from the GSW fold-5 model. Thirteen monthly checkpoints use data available before each month's first scored origin. The three-month yield maps to a price proxy:
\begin{equation}
 P_t=100\exp\{-0.25\,y_t/100\},\qquad L_{t+1}=P_t-P_{t+1}.
\end{equation}
Maturity stays at 0.25 years, and quoted yields are treated as continuously compounded. This proxy omits Treasury-bill quotation conventions, maturity roll-down, and cash flows.

The 252 forecast origins run from April 29, 2025 to April 30, 2026, each paired with the next available observation. The exercise compares the frozen checkpoint with monthly refitting. It overlaps the final main-study fold and was inspected during development.

\begin{table}[htbp]
\centering\small
\caption{Saved VaR diagnostics for the stabilized model. ``UC'' denotes the Kupiec unconditional-coverage test; ``Ind.'' and ``CC'' denote Christoffersen independence and conditional-coverage tests. All reported values after the breach count are $p$-values.}
\begin{tabular}{llrrrr}
\toprule Updating & Level & Breaches / 252 & UC & Ind. & CC\\\midrule
Frozen & 95\% & 12 & 0.861 & 0.013 & 0.044\\
Frozen & 99\% & 6 & 0.061 & 0.588 & 0.150\\
Monthly & 95\% & 16 & 0.345 & 0.983 & 0.640\\
Monthly & 99\% & 6 & 0.061 & 0.588 & 0.150\\\bottomrule
\end{tabular}
\label{tab:risk}
\end{table}

At 95\%, the frozen model's breach rate is close to 5\%, but independence and conditional coverage are rejected at the 5\% level \citep{kupiec,christoffersen}. Monthly refitting increases breaches from 12 to 16 and reduces clustering. All reported monthly-refit tests have $p>0.05$.

Both 99\% rows have six breaches and transition counts $(n_{00},n_{01},n_{10},n_{11})=(239,6,6,0)$. Daily losses and forecasts reproduce all three $p$-values. Frozen-model breach origins are May 1, May 6, May 9, May 30, and October 31, 2025, and February 27, 2026. Monthly-refit breach origins are April 30, May 9, May 30, October 31, December 23, and December 26, 2025. Equal counts explain the identical statistics despite different breach dates. Six exceedances offer little information about extreme risk.

The Expected Shortfall diagnostic bootstraps centered exceedance residuals in a one-sided test inspired by \citet{mcneil}. Frozen-model $p$-values are 0.0073 (95\%) and 0.0342 (99\%), both rejecting at 5\%. Monthly-refit values are 0.0753 and 0.0632, respectively. Only the monthly-refit rows are not rejected at 5\%. Their mean residuals remain positive. The bootstrap resamples few exceedances independently without fully accounting for serial dependence or conditional heteroskedasticity. Non-rejection does not validate Expected Shortfall.

\end{document}